\documentclass[showpacs,amssymb,floatfix]{revtex4}
\usepackage{graphicx}
\usepackage{dcolumn}
\usepackage{bm}
\begin{document}
\title{Frequency comb generation for wave transmission through the nonlinear dimer}
\author{Konstantin N. Pichugin and Almas F. Sadreev}
\affiliation{Institute of Physics, 660036, Krasnoyarsk, Russia}
\date{\today}
\begin{abstract}
We study dynamical response of a nonlinear dimer to a symmetrically injected monochromatic wave.
We find a domain in the space of frequency and amplitude of the injected wave where all stationary
solutions are unstable.
In this domain  scattered waves
carry multiple harmonics with equidistantly spaced frequencies (frequency comb effect).
The instability is related to a symmetry
protected bound state in the continuum whose response
is singular as the amplitude of the injected wave tends to zero.
\end{abstract}
\maketitle
\section{Introduction}
A nonlinear quantum dimer represents the simplest
realization of the discrete nonlinear Schr\"odinger
equation and has attracted much interest for a long time
\cite{Eilbeck,Campbell,Tsironis,Kenkre0,Kenkre,Tsironis1,Molina,Miros,Pickton,Xu}.
The interest is related to the phenomenon of symmetry breaking (self-trapping)
\cite{Eilbeck,Campbell,Molina,Berstein}. On the other hand, non-trivial time-dependent solutions
were found in the nonlinear dimer \cite{Eilbeck,Campbell,Tsironis,Kenkre0,Kenkre,Tsironis1,Pickton}.

The observation of these remarkable properties
of the closed nonlinear dimer implies application of a probing wave that opens
the dimer. Respectively, temporal equations describing the open nonlinear dimer
become non-integrable which constitutes the main difference between
the closed and open nonlinear dimers. As dependent on the way of opening
the transmission through nonlinear dimer was studied in
Refs. \cite{Maes1,Maes2,Miros2,BPS,Brazhnyi,Miros3,Shapira,T,Maksimov,Xu-Miros,Barashenkov}
where the phenomenon of symmetry breaking was reported.
What is interesting there is a domain in the space of frequency and
amplitude of injected wave where stable stationary solutions of
the temporal equations do not exist \cite{BPS,Maksimov}. Thus one can expect that the dynamical
response of the nonlinear dimer will display features which can not be described by the
stationary scattering theory. In particular injection of a monochromatic symmetric wave into the
nonlinear plaquette gives rise to emission of anti-symmetric satellite waves with frequencies different from
the frequency of the incident wave \cite{Maksimov}.
This phenomenon is known as frequency comb (FC) generation and is widely studied in various linear and
nonlinear systems \cite{Science,Lushui,Abrams}.
The FC generation can be interpreted as a modulational
instability of the continuous-wave pump mode \cite{Matsko} revealed even in a single off-channel
nonlinear cavity \cite{Hansson}. In this paper we show that
illumination of the nonlinear dimer with a monochromatic wave
in the domain of instability gives rise to FC generation.

\section{Coupled mode theory equations}
\begin{figure}[ht]
\includegraphics[height=7cm]{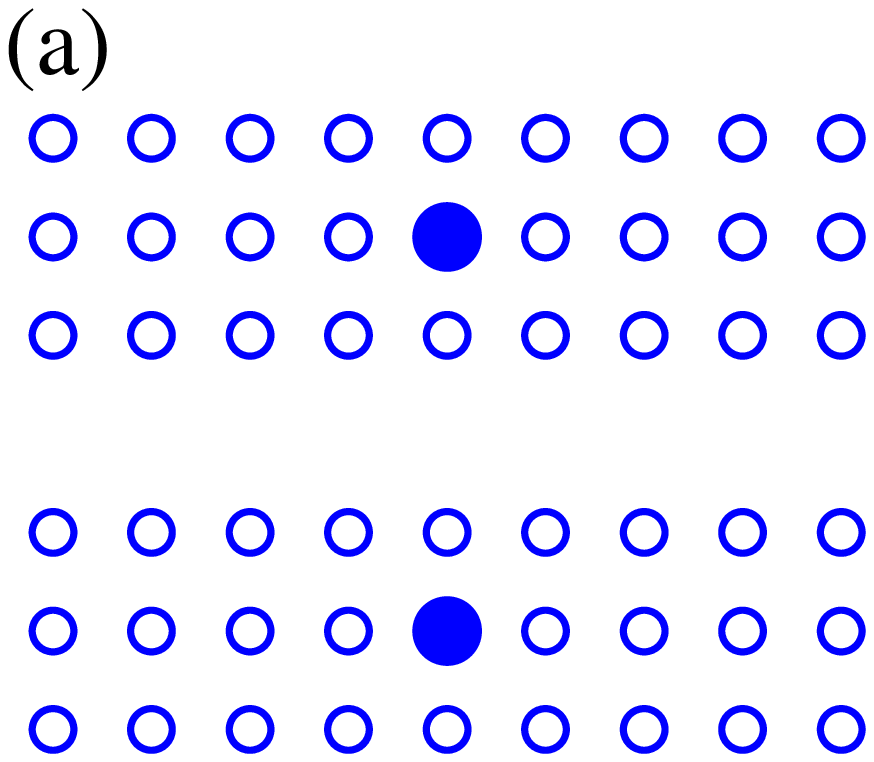}
\includegraphics[height=7cm]{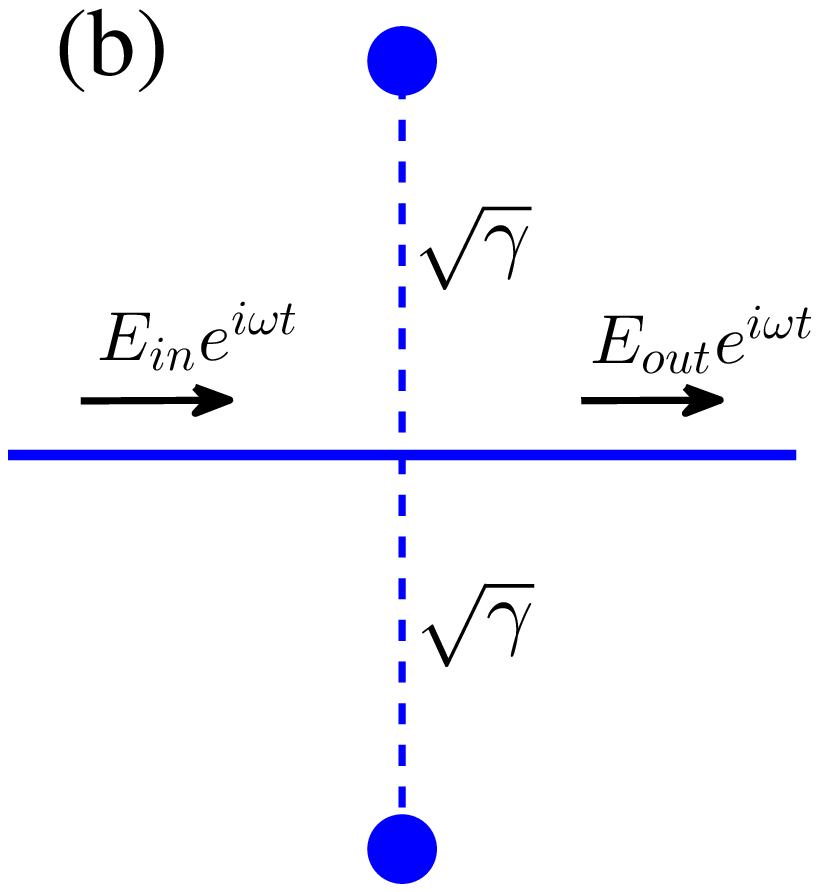}
\caption{(a) Two identical microcavities made from a Kerr media marked by
filled circles are inserted into the square lattice photonic
crystal of dielectric rods. The 1D waveguide is formed by
extraction  of a linear chain. (b) Two nonlinear sites
marked by filled circles are positioned symmetrically relative to waveguide and form open nonlinear dimer.}
\label{fig1}
\end{figure}

The nonlinear dimer can be realized in 2D  photonic crystal in the form of coupled microcavities
based on two identical dielectric defect rods
with the Kerr effect \cite{BPS} as shown in Fig. \ref{fig1} (a).
Taking the radii of the rods small enough we can present each rod by a single site
variable $A_j, j=1, 2$ disregarding space inhomogeneity of electromagnetic field in the rods.
In terms of the eigenfunctions of the Maxwell equations for each microcavity it means that only the monopole mode
with the eigenfrequency $\omega_0$ resides
in the photonic crystal waveguide propagation band and thereby is relevant in the scattering. For simplicity
we disregard the
dispersion properties of the waveguide and  write for the dimer illuminated by light with
the amplitude $E_{in}$ and frequency $\omega$ the following temporal coupled mode theory (CMT) equations \cite{Haus,Suh,BPS}
\begin{eqnarray}\label{CMTtemp}
&-i\dot{A_1}=(\omega_0+\lambda|A_1|^2)A_1+uA_2+i\gamma(A_1+A_2)-i\sqrt{\gamma}E_{in}e^{i\omega t},&\nonumber\\
&-i\dot{A_2}=(\omega_0+\lambda|A_2|^2)A_2+uA_1+i\gamma(A_1+A_2)-i\sqrt{\gamma}E_{in}e^{i\omega t}.&
\end{eqnarray}
Here the terms $\lambda|A_j|^2A_j, j=1, 2$ account for the Kerr effect of each microcavity,
the term $\sqrt{\gamma}$ is responsible for the coupling of the off-channel  cavity with
the waveguide. The monopole mode of each cavity is localized within a few lattice units \cite{Joanbook}. If
the cavities are positioned far from waveguide we can neglect direct coupling $u$ between them.
However even for this case the open nonlinear dimer remains cardinally different
from the case of the closed dimer because of the interaction between the cavities via the continuum.
For simplicity we consider the case $u=0$.
The amplitude of the transmitted  wave is given by the following equation \cite{Suh,BPS}
\begin{equation}\label{t}
E_{out}=E_{in}-\sqrt{\gamma}(a_1+a_2)
\end{equation}
where $A_j(t)=a_j(t)e^{i\omega t}$.
The open dimer governed by the CMT equations (\ref{CMTtemp}) is shown in Fig. \ref{fig1} (b).

Assuming that the solution is stationary $a_j(t)=a_{j0}=const$ the CMT equations (\ref{CMTtemp})
are simplified to
\begin{eqnarray}\label{CMTstat}
&(\nu+\lambda|a_{10}|^2)a_{10}+i\gamma(a_{10}+a_{20})-i\sqrt{\gamma}E_{in}=0\,&\nonumber\\
&(\nu+\lambda|a_{20}|^2)a_{20}+i\gamma(a_{10}+a_{20})-i\sqrt{\gamma}E_{in}=0\,&
\end{eqnarray}
where $\nu=\omega_0-\omega$. As the input amplitude
$E_{in}$ increases  the system bifurcates from the symmetry preserving solution $a_{10}=a_{20}$ to
the symmetry breaking solution $I_1\neq I_2, \theta_1-\theta_2=0, \pi$ where
$a_{j0}=\sqrt{I_j}\exp(i\theta_j), j=1,2$
similar to the closed nonlinear dimer \cite{Eilbeck,Tsironis,Kenkre}.
However in contrast to the closed nonlinear dimer the open dimer can also transit into the phase
symmetry breaking solution with $I_1=I_2$ but $\theta_1-\theta_2\neq 0, \pi$ \cite{BPS,Maksimov}.


\section{instability of stationary solutions}
Numerical analysis of stability of the stationary solutions revealed a domain in the space of parameters
$\omega$ and $E_{in}$ where {\it all} stationary solutions are unstable \cite{BPS}.
Similar result was found in the open plaquette
of four nonlinear cites \cite{Maksimov}. In this section we find the domain
of instability of stationary solutions of temporal equations (\ref{CMTtemp}) analytically.

To examine the stability of solutions of Eq. (\ref{CMTtemp}) we apply a standard small perturbation technique
\cite{Litchinitser,Cowan}:
\begin{equation}\label{small}
    a_j(t)=a_{j0}+(x_j+iy_j)e^{\mu t}, ~~j=1,2,
\end{equation}
where the second term in Eq. (\ref{small}) is considered to be small.
For the symmetry preserving solutions we have from Eq. (\ref{CMTstat})
\begin{equation}\label{A10}
    a_{10}=a_{20}=A_0=\frac{i\sqrt{\gamma}E_{in}}{\nu+\lambda I_0+2i\gamma}
\end{equation}
 where according to Eq. (\ref{CMTtemp})
\begin{equation}\label{I}
    I_0[(\nu+\lambda I_0)^2+4\gamma^2]=\gamma E_{in}^2,
\end{equation}
and $I_0=|A_0|^2$.
Substituting (\ref{small}) into Eq. (\ref{CMTtemp}) we obtain
the following system of algebraic equations
\begin{eqnarray}\label{x-y}
&-(\mu+2\lambda Re(A_0)Im(A_0))(x_1-x_2)=(\nu+\lambda I_0+2\lambda Im(A_0)^2)(y_1-y_2),&\nonumber\\
&(\mu-2\lambda Re(A_0)Im(A_0))(y_1-y_2)=(\nu+\lambda I_0+2\lambda Re(A_0)^2)(x_1-x_2),&
\end{eqnarray}
and
\begin{eqnarray}\label{x+y}
&-(\mu+2\gamma+2\lambda Re(A_0)Im(A_0))(x_1+x_2)=(\nu+\lambda I_0+2\lambda Im(A_0)^2(y_1+y_2)&\nonumber\\
&(\mu+2\gamma-2\lambda Re(A_0)Im(A_0))(y_1+y_2)=(\nu+\lambda I_0+2\lambda Re(A_0)^2)(x_1+x_2).&
\end{eqnarray}
For Eq. (\ref{x-y}) we obtain  the eigenvalues
\begin{equation}\label{mu}
    \mu^2=-(\nu+\lambda I_0)(\nu+3\lambda I_0).
\end{equation}
The equation $\mu=0$ defines
the boundary where the symmetry preserving family of the stationary solutions becomes unstable \cite{Cowan}.
From Eq. (\ref{mu}) we  obtain equations $\nu+\lambda I_c=0$ and $\nu+3\lambda I_c=0$.
Substituting these values of $I_c$ into  Eq. (\ref{I}) we obtain for the boundaries of the domain where
stable  stationary solutions do not exist
\begin{eqnarray}\label{domain}
&E_{in}^2=-\frac{4\gamma\nu}{\lambda},&\nonumber\\
&E_{in}^2=-\frac{4\nu}{3\lambda\gamma}[\nu^2/9+\gamma^2].&
\end{eqnarray}

\begin{figure}
\includegraphics[height=8cm,width=8cm,clip=]{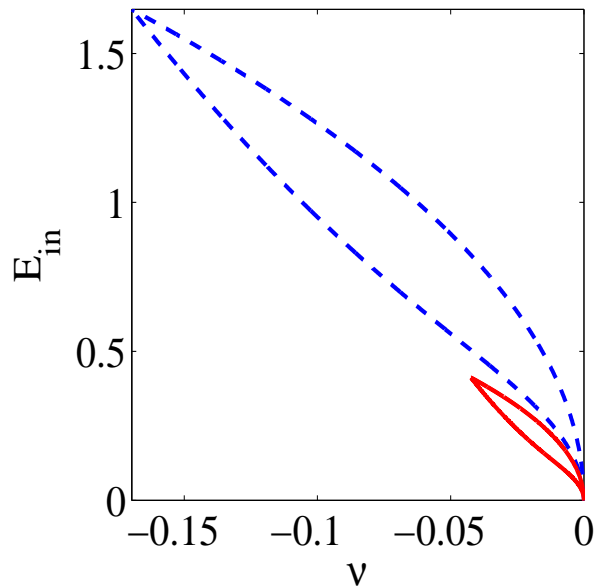}
\caption{(Color online) Domain in space of physical parameters $\nu=\omega_0-\omega$ and
$E_{in}$ where all stationary solutions are unstable for $\gamma=0.01$
(red solid line) and $\gamma=0.04$ (blue dash line). Other parameters are
$\omega_0=1, \lambda=0.01$.} \label{fig2}
\end{figure}
The domains of instability of stationary solutions are shown in  Fig. \ref{fig2}. Eq. (\ref{x+y})
does not give  contribution into the domain of instability. Also numerical analysis of the stability has shown
that the symmetry breaking and phase symmetry breaking solutions fall into the same domain of stability
as the symmetry preserving solution.
\section{numerical solutions of nonlinear temporal CMT equations}
\begin{figure}
\includegraphics[height=8cm]{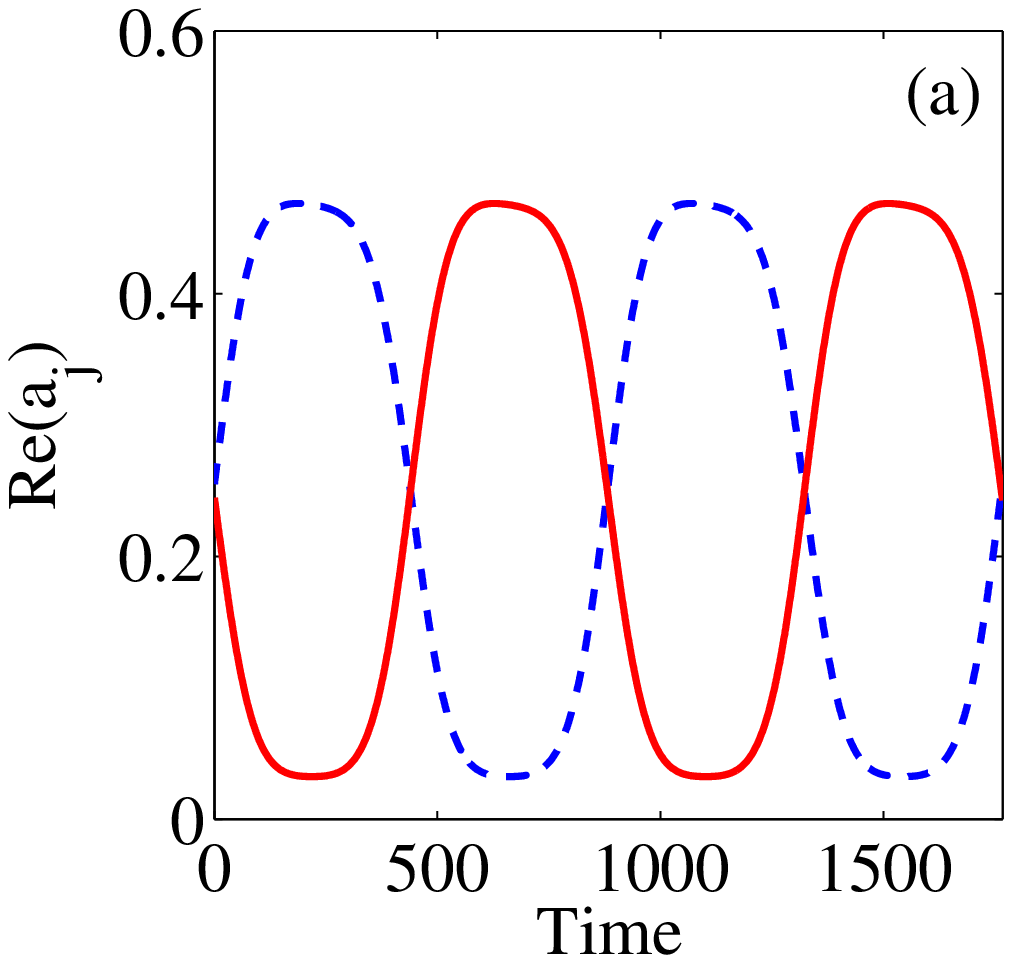}\hfill
\includegraphics[height=8cm]{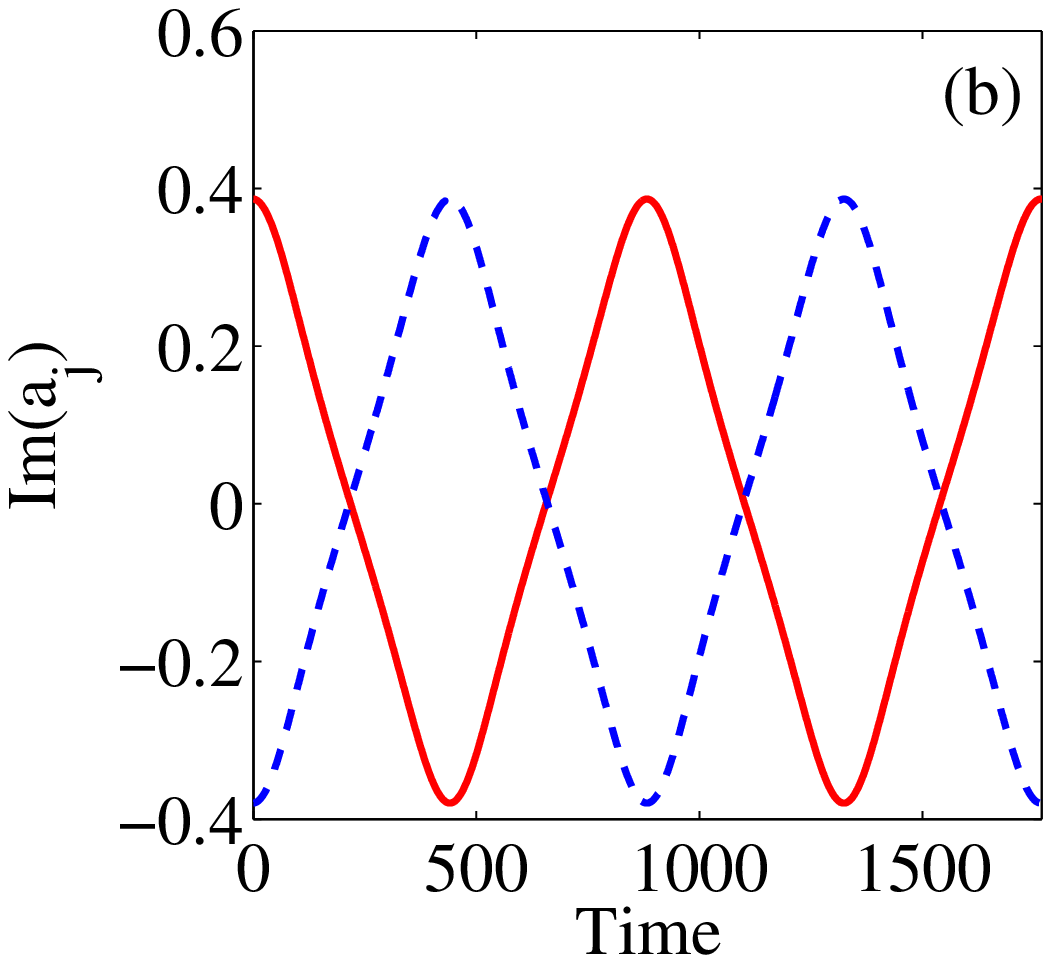}
\caption{(Color online) Time evolution of site amplitudes $A_j(t), j=1, 2$,
real and imaginary parts for $E_{in}=0.1, \nu=-0.001, \gamma=0.04, \lambda=0.01$.}
\label{fig3}
\end{figure}

\begin{figure}
\includegraphics[height=8cm]{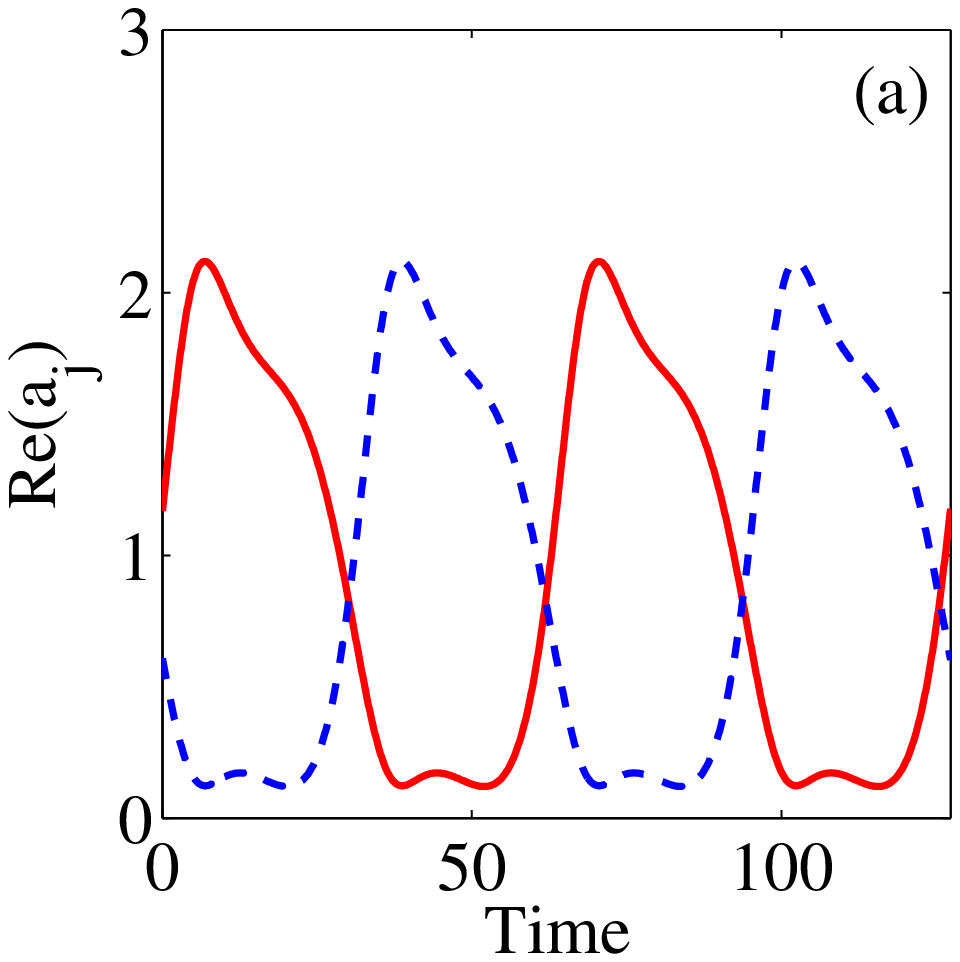}\hfill
\includegraphics[height=8cm]{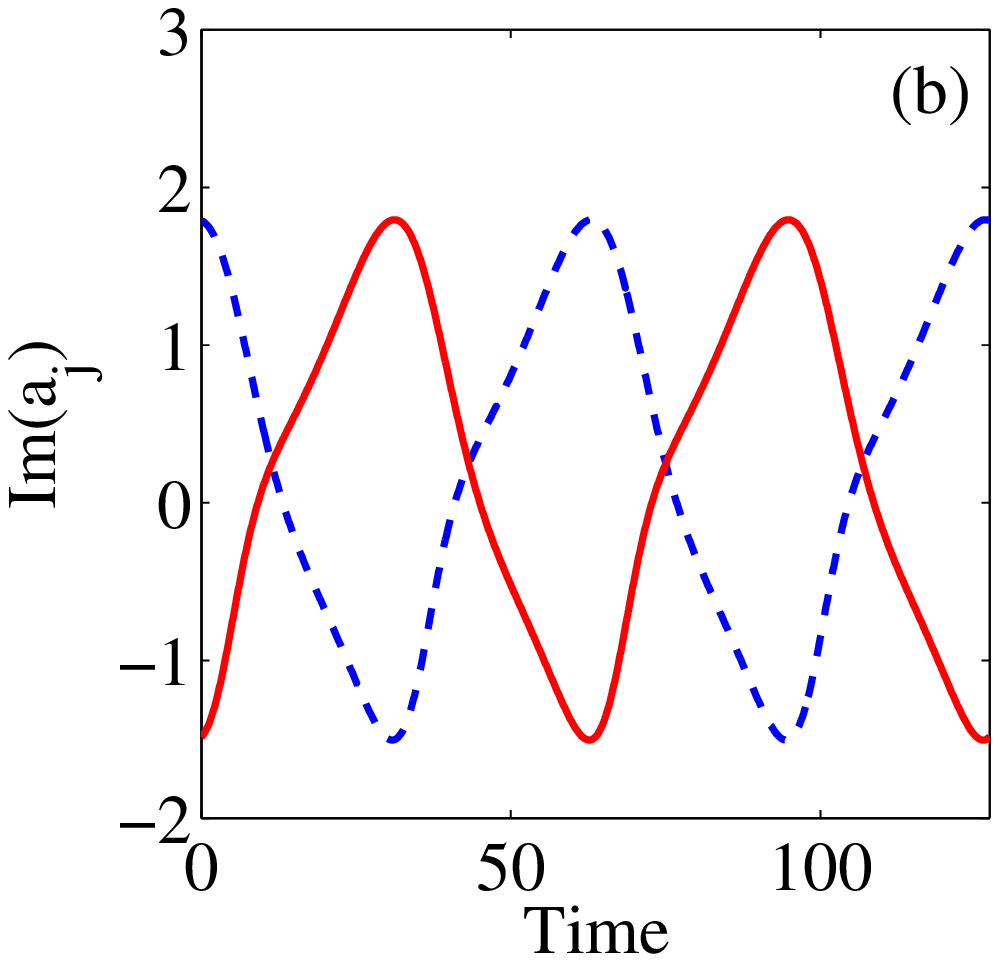}
\caption{(Color online) The same as in Fig. \ref{fig3} but for the parameters
$E_{in}=0.4, \nu=-0.02, \gamma=0.04, \lambda=0.01$.}
\label{fig4}
\end{figure}

Substituting $A_j(t)=a_j(t)\exp(i\omega t)$ into the temporal CMT equations
(\ref{CMTtemp}) we obtain
\begin{eqnarray}\label{CMTtem}
&-i\dot{a_1}=(\nu+\lambda|a_1|^2)a_1+i\gamma(a_1+a_2)-i\sqrt{\gamma}E_{in},&\nonumber\\
&-i\dot{a_2}=(\nu+\lambda|a_2|^2)a_2+i\gamma(a_1+a_2)-i\sqrt{\gamma}E_{in}.&
\end{eqnarray}
One can see that the solutions possess a symmetry with the half period time shift corresponding to the
permutation of the sites
\begin{equation}\label{sym1}
a_1(t+T/2)=a_2(t), a_2(t+T/2)=a_1(t).
\end{equation}
Indeed, after time shift $t\rightarrow t+T/2$ in the first equation in (\ref{CMTtem})
we obtain the second equation using Eq. (\ref{sym1}) and the periodicity of
the solutions. Thus, the system of equations (\ref{CMTtem}) is reduced to
one temporal equation
\begin{equation}\label{sym2}
-i\dot{a_j}=(\nu+\lambda|a_j(t)|^2)a_j(t)+i\gamma(a_j(t)+a_j(t+T/2))-i\sqrt{\gamma}E_{in}.
\end{equation}
Nevertheless the symmetry (\ref{sym1}) does not allow to solve  Eq. (\ref{sym2})
because of unknown period $T$ which strongly depends on the
intensity of the injected wave. In Fig. \ref{fig3} and Fig. \ref{fig4}
we present the results of  numerical simulations
of Eq. (\ref{CMTtem}) in the domain of unstable stationary solutions which
demonstrate the symmetry (\ref{sym1}). Fig. \ref{fig4} also demonstrates rachet effect due to
absence of the time reversal symmetry in the open dimer. We chose the parameters listed in caption of
Fig. \ref{fig3} and Fig. \ref{fig4} guided by the data on photonic crystal
microcavities from Ref. \cite{BPS}.

\begin{figure}
\includegraphics[height=8cm]{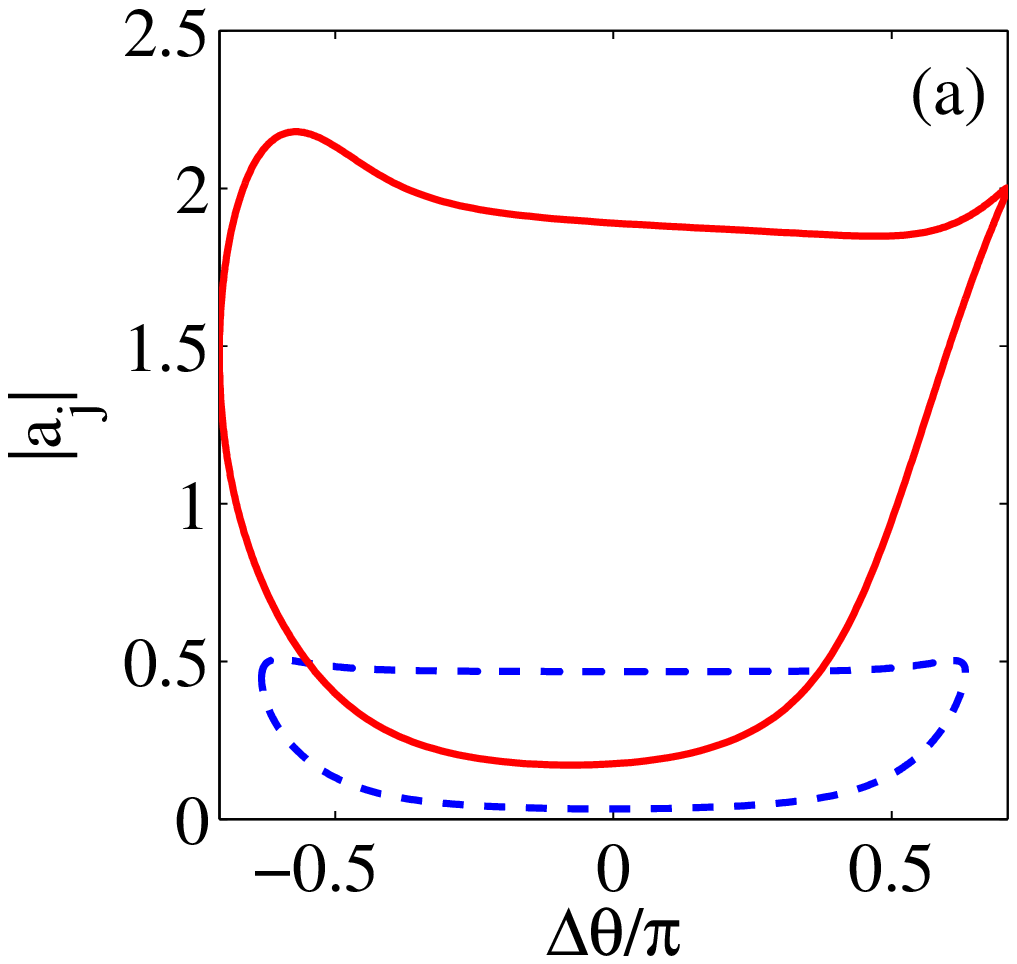}\hfill
\includegraphics[height=8cm]{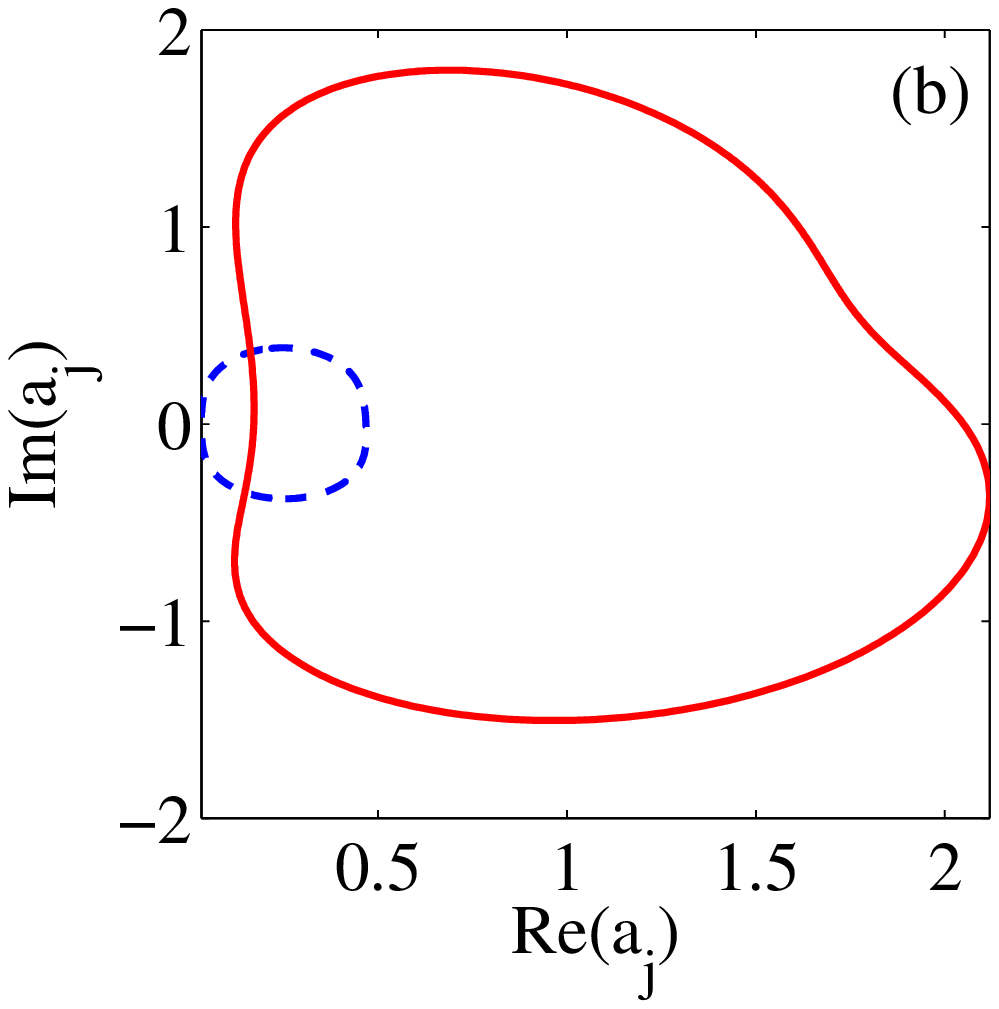}
\caption{(Color online) (a) Trajectories projected onto $|a_j|$ and phase difference
$\Delta\theta=\theta_1-\theta_2$  and (b) real and imaginary parts of amplitudes $a_j(t), j=1,2$
for different points in the domain of instability: $E_{in}=0.1, \nu=-0.001$ (blue dash line) and
$E_{in}=0.4, \nu=-0.02$ (red solid line). Other parameters are $\omega_0=1, \gamma=0.04, \lambda=0.01$.}
\label{fig5}
\end{figure}

In order to compare the results with the closed dimer
Ref. \cite{Eilbeck} we present trajectories projected onto the modulus $|a_j|$ and phase difference $\Delta\theta$
between cavities in Fig. \ref{fig5} (a). Although for a small injected amplitude $E_{in}$ the trajectories
look similar to those shown in Ref. \cite{Eilbeck} however  with the growth of
$E_{in}$ the trajectories become asymmetrical relative to $\Delta\theta\rightarrow -\Delta\theta$.
The trajectories
projected onto the real and imaginary parts of the amplitudes $a_j(t)$
demonstrate the most striking difference
between the closed and open nonlinear dimer as shown in Fig. \ref{fig5} (b). While for the closed dimer
the trajectories form circles centered at the origin of the coordinate system (they are not shown in
Fig. \ref{fig5} (b)) the trajectories of the open dimer are shifted relative to the coordinate origin.
Phase transformation of the
injected wave $E_{in}\rightarrow E_{in}e^{i\alpha}$ rotates the trajectories in Fig. \ref{fig5} (b) by
the same angle $\alpha$.

\begin{figure}
\includegraphics[height=7cm,width=7cm,clip=]{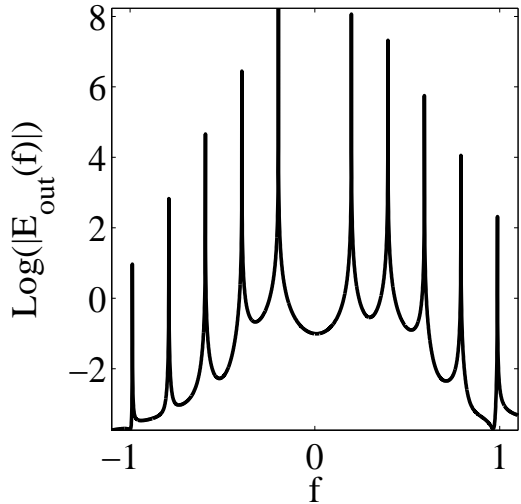}
\caption{Fourier transform $E_{out}(f)$  of the transmitted wave
in Log scale for the parameters listed in Fig. \ref{fig4}.} \label{fig6}
\end{figure}

It is clear that such a complicated time behavior of the amplitudes will reflect at the
transmitted wave according to Eq. (\ref{t}). Fig. \ref{fig6}  shows the Fourier transformation
of the transmitted wave
\begin{equation}\label{trans}
    E_{out}(t)=\int df E_{out}(f)\exp(ift)
\end{equation}
that demonstrates sharp peaks spaced equidistantly, FC comb effect. In what follows we define
the interval between the peaks of the Furrier transform $F_{out}(f)$ as the FC period $\Omega_{FC}$.
One can see from Fig. \ref{fig6} $|E_{out}(f)|\neq |E_{out}(-f)|$
that is a consequence of the rachet effect as seen from Fig. \ref{fig4}.
\section{the asymptotic evaluation of the frequency comb period}
The reason for the cardinal difference between the closed and open nonlinear dimers is the symmetry of the
system. Let us rewrite Eq. (\ref{CMTtem})
in terms of the eigenmodes of the closed linear dimer
\begin{eqnarray}\label{asaa}
&-i\dot{a}_s=(\nu+\lambda[|a_s|^2+2|a_a|^2])a_s+\lambda a_a^2a_s^{*}+2i\gamma a_s-i\sqrt{\gamma}
E_{in},&\nonumber\\
&-i\dot{a}_a=(\nu+\lambda[|a_a|^2+2|a_s|^2])a_a+\lambda a_s^2a_a^{*}.&
\end{eqnarray}
\begin{figure}
\includegraphics[height=7cm,width=7cm,clip=]{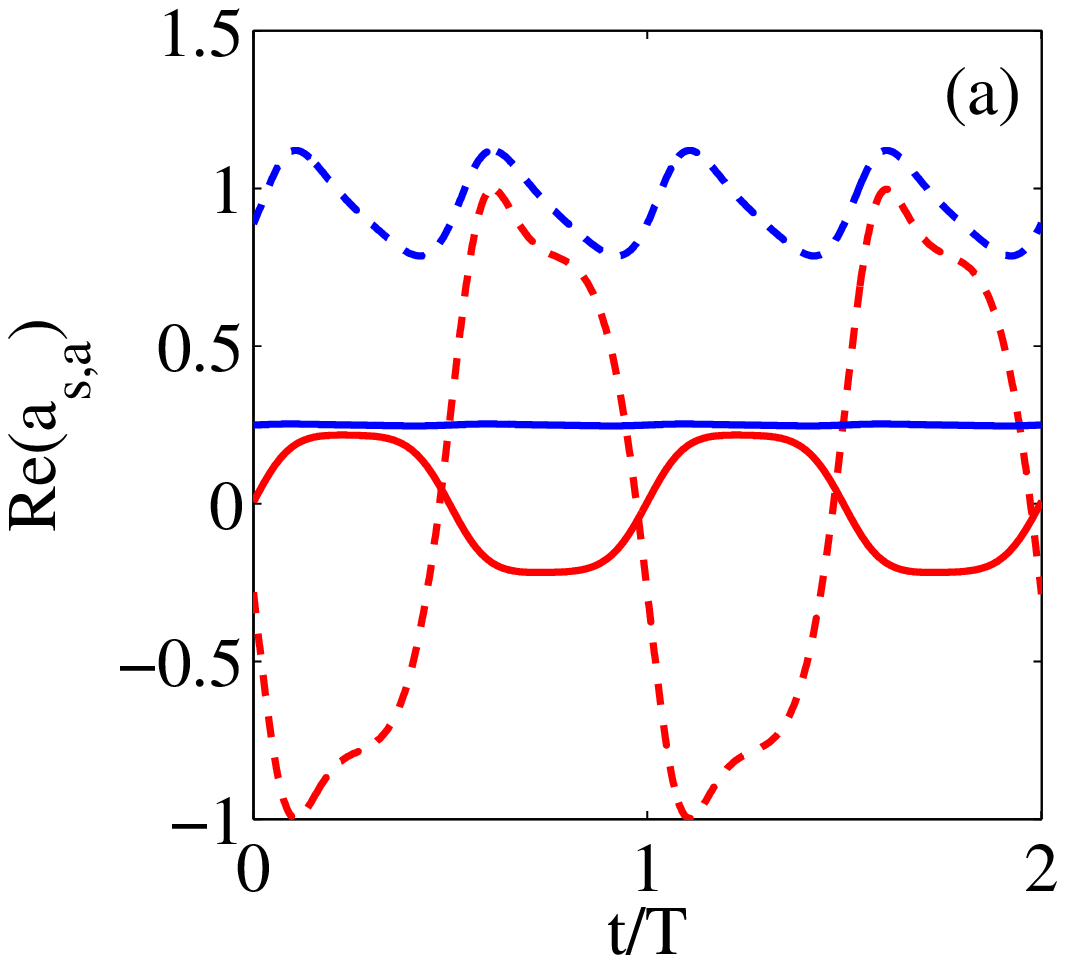}
\includegraphics[height=7.5cm,width=7.5cm,clip=]{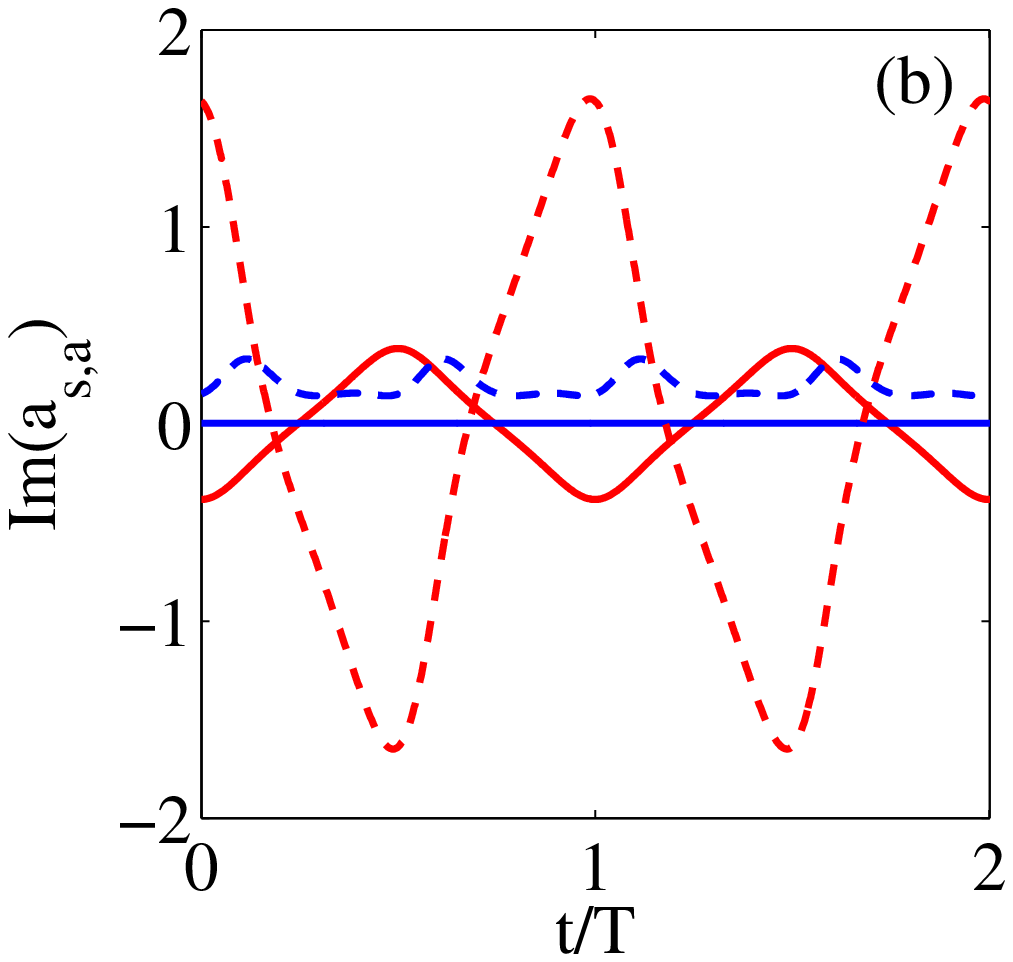}
\caption{(Color online) Time evolution of real parts (a) and imaginary parts (b)
of symmetric $a_s=\frac{1}{2}(a_1+a_2)$ (blue lines) and antisymmetric
mode $a_a=\frac{1}{2}(a_1-a_2)$ (red lines) for  $E_{in}=0.1, \nu=-0.001$ (solid lines) and
$E_{in}=0.2, \nu=-0.02$ (dash lines). Other parameters are $\omega_0=1, \gamma=0.04, \lambda=0.01$.} \label{fig7}
\end{figure}
where $a_{s,a}=\frac{1}{2}(a_1\pm a_2)$ are the symmetric and antisymmetric eigenmodes of the dimer
with the eigenfrequencies $\omega_{s,a}=\omega_0$.

Let us take temporarily the dimer linear.
The design of the open dimer (Fig. \ref{fig1}) implies that the injected wave can probe only the
symmetric mode with a Breit-Wigner response $a_s=i\sqrt{\gamma}E_{in}/(\nu+2i\gamma)$,
while the antisymmetric mode remains hidden
as seen from the CMT equations (\ref{asaa}). It oscillates with the frequency $\nu$ however with
the uncertain amplitude. That defines the antisymmetric mode $a_a$
as a symmetry protected bound state in the continuum \cite{photonics,Moiseyev,Segev,Wei}.
Returning to the site amplitudes we therefore obtain
\begin{equation}\label{a1a2}
    a_j=\frac{i\sqrt{\gamma}E_{in}}{\nu+2i\gamma}\pm ae^{i\nu t},~~j=1,2
\end{equation}
making the time behavior of the site amplitudes of the linear dimer non stationary. This equation
constitutes the time dependent contribution of the bound state in the continuum established for the
stationary case in quantum mechanical \cite{ring} and photonic crystal systems \cite{photonics}.

The nonlinearity results in two effects.
The first obvious result is that
the resonance eigenfrequency $\nu+\lambda I_0$ of the symmetric mode is shifted proportional
to $E_{in}^2$. That agrees with behavior of the instability domain at small $E_{in}$
as derived in Section III and shown in Fig. \ref{fig2}. The second effect is more sophisticated.
For small $E_{in}$ the symmetric mode $a_s$ is almost constant while oscillations
of the antisymmetric mode $a_a$ are dominant as shown in Fig. \ref{fig7}. As seen from
the first equation in Eq. (\ref{asaa}) the antisymmetric mode plays the role of a driving force for the
mode $a_s$ via the the nonlinear term $\lambda a_a^2a_s^{*}$. Then if the frequency of the
mode $a_a$ is $\Omega$ then the symmetric mode oscillates with double frequency $2\Omega$
as it is seen from the numerical solution in Fig. \ref{fig7}. Respectively, the transmitted wave
carries the harmonics with the same frequency $2\Omega$ in accordance to Eq. (\ref{t}).

In order to consider these nonlinear effects in the open nonlinear dimer we use the asymptotic
methods by Bogoliubov and Mitropolsky \cite{Bogol}. Eq. (\ref{asaa}) can be rewritten as follows
\begin{eqnarray}\label{asaa1}
&i\dot{a}_s+(\nu+2i\gamma)a_s-i\sqrt{\gamma}E_{in}=\varepsilon F_s(a_s,a_a),&\nonumber\\
&i\dot{a}_a+\nu a_a=\varepsilon F_a(a_s,a_a)&
\end{eqnarray}
where the parameter $\lambda$ is considered as a small parameter $\varepsilon$ and functions $F_{s,a}$ are
polynomial functions of $a_{s,a}$ determined by Eq. (\ref{asaa}).
Then the solution up to the first order in $\varepsilon$ can be sought in the form
\begin{eqnarray}\label{sol}
&a_s=s_0(a,\phi)+\varepsilon s_1(a,\phi)&\nonumber\\
&a_a=a_0(a,\phi)+\varepsilon a_1(a,\phi)&
\end{eqnarray}
as functions of the amplitude $a$ and phase $\phi$. They are given by the following equations
\begin{eqnarray}\label{deriv}
&\dot{a}=\varepsilon D_1(a)&\nonumber\\
&\dot{\phi}=\nu+\varepsilon \Omega_1(a)&
\end{eqnarray}
where $\nu$ is the frequency of oscillations at $\varepsilon=0$.
Substitution of Eqs. (\ref{sol}), (\ref{deriv}) and relation
\begin{equation}\label{deriv_sol}
\dot{a}_{s,a}=\dot{a}\frac{\partial a_{s,a}}{\partial a}+\dot{\phi}\frac{\partial a_{s,a}}
{\partial \phi}.
\end{equation}
into Eq. (\ref{asaa}) gives the following equation at the zeroth order in parameter $\varepsilon$
\begin{equation}
s_0(a,\phi)=\frac{i\sqrt{\gamma}E_{in}}{\nu+2i\gamma}, ~~a_0(a,\phi)=a\exp(i\phi),
\end{equation}
where the amplitude $a$ is undefined.

\begin{figure}
\includegraphics[height=8cm,width=8cm,clip=]{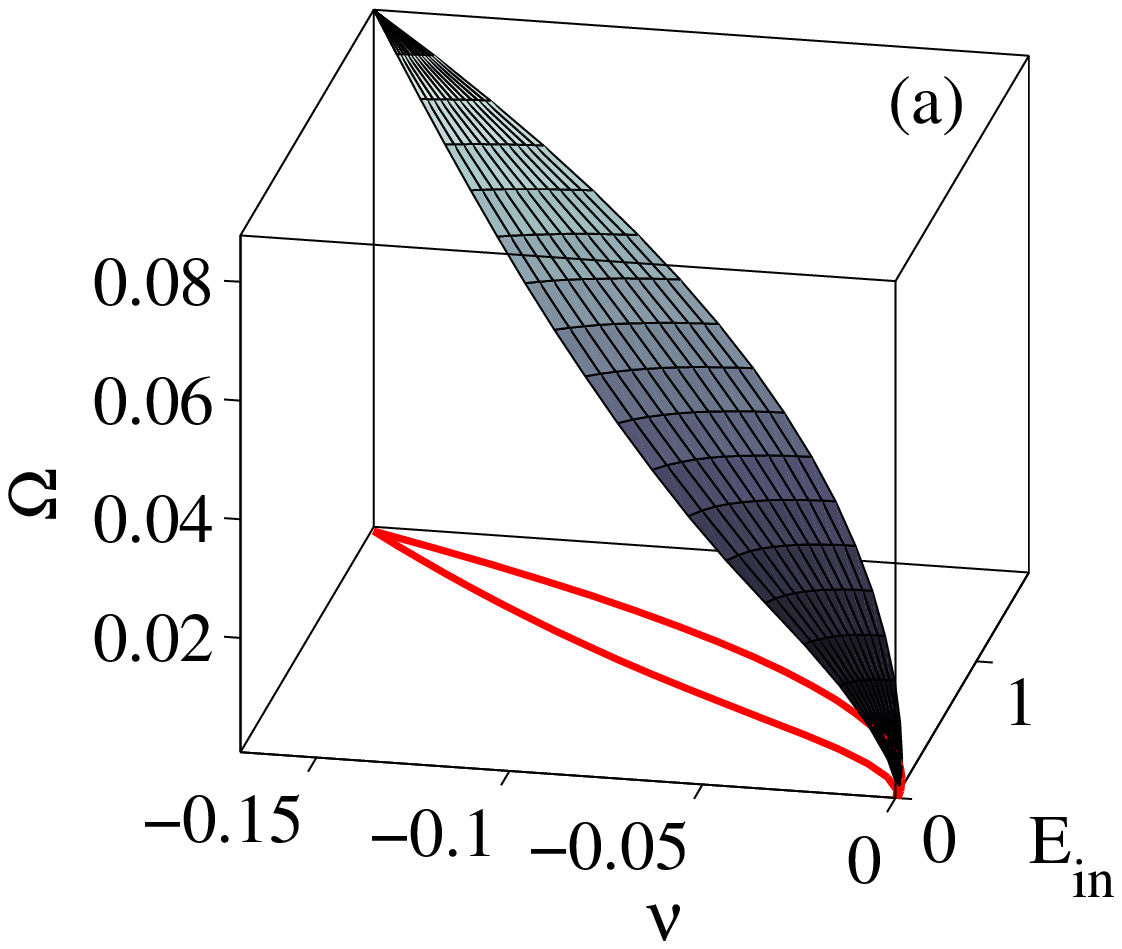}\hfill
\includegraphics[height=8cm,width=8cm,clip=]{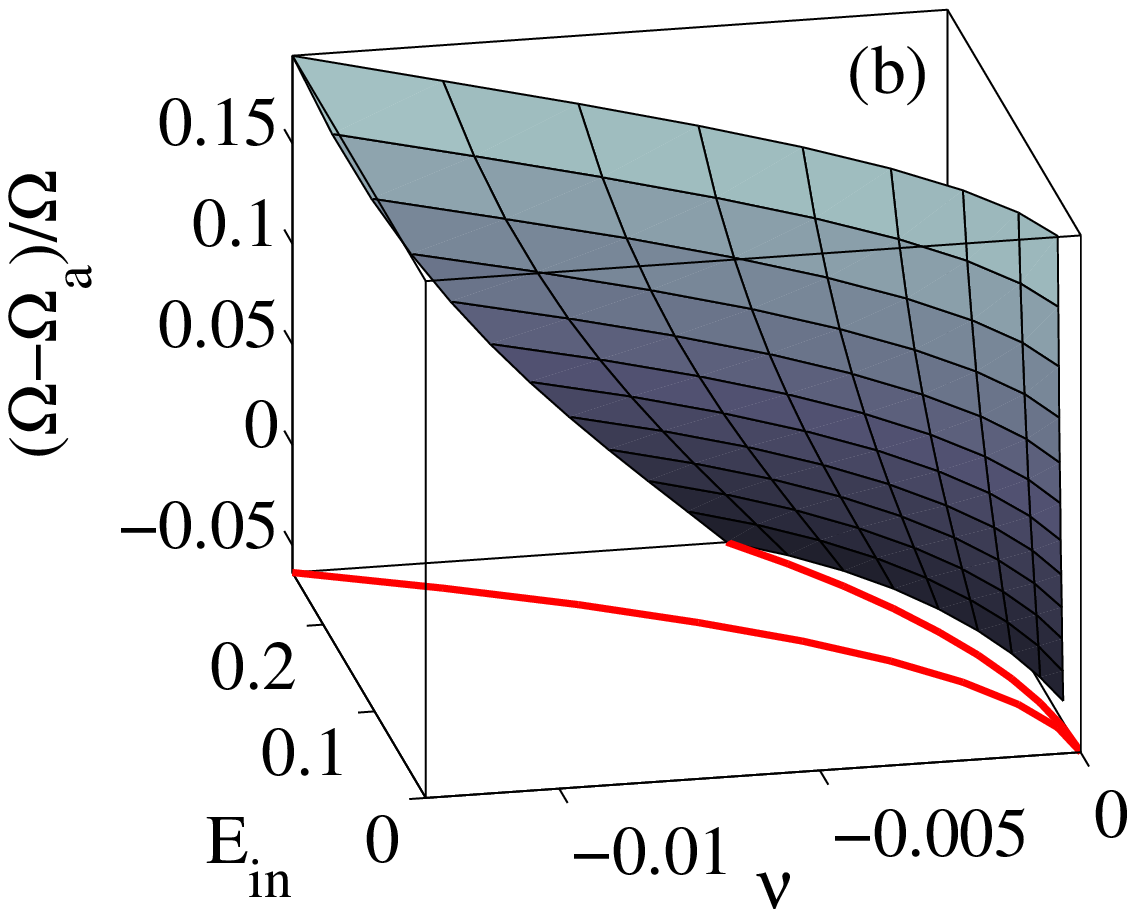}
\caption{(Color online) (a) Period of harmonics $\Omega$ vs amplitude $E_{in}$
and frequency $\nu$ of injected monochromatic wave calculated numerically. (b) Difference
between numerical data and analytical results given by Eq. (\ref{Omega}). The parameters of the dimer are
$\omega_0=1, \gamma=0.01, \lambda=0.01$. Below the domain of
instability defined by Eq. (\ref{domain}) is shown by red lines.}
\label{fig8}
\end{figure}

In the first order in  $\varepsilon$ we obtain the following equations
\begin{eqnarray}\label{first}
&-i\nu\frac{\partial s_1}{\partial \phi}=(\nu+2i\gamma)s_1+a_0^2s_0^*+(|s_0|^2+2a^2)s_0&\nonumber\\
&-i\nu\frac{\partial a_1}{\partial \phi}=\nu a_1+s_0^2a_0^*+(a^2+2|s_0|^2)a_0+(iD_1-a\Omega_1)
\exp(i\phi).&
\end{eqnarray}
One can expand
\begin{eqnarray}\label{s1a1}
&s_1(a,\phi)=\sum_n F_{s,n}(a) \exp(in\phi)&\nonumber\\
&a_1(a,\phi)=\sum_n F_{a,n}(a) \exp(in\phi)&.
\end{eqnarray}
According to Ref. \cite{Bogol} there is an uncertainty in choice of functions
$s_1$ and $a_1$ that allows to exclude, for example, the first harmonic contributions
$F_{s,1}, F_{a,1}$ that gives the following equations
\begin{equation}
D_1(a)=0, ~~\Omega_1(a)=a^2+2|s_0|^2.
\end{equation}
Then, solutions of (\ref{first}) are the following
\begin{eqnarray}\label{sol_1}
&s_1(a,\phi)=-\frac{|s_0|^2+2a^2}{\nu+2i\gamma}s_0+\frac{a^2}{\nu-2i\gamma}s_0^*\exp(2i\phi)&\nonumber\\
&a_1(a,\phi)=-\frac{a}{2\nu}s_0^2\exp(-i\phi).&
\end{eqnarray}
This equations show that the symmetric solution consists of even terms $n=0, \pm 2,\ldots$ in the expansion
(\ref{s1a1}) while the antisymmetric solution
consists of the odd terms $n=\pm 1, \pm 3, \ldots$. The higher orders in the small parameter holds these features.
From Eq. (\ref{deriv}) we have
\begin{equation} \label{Omega}
\phi=\left(\nu+\lambda \left(a^2+2|s_0|^2\right)\right)t=\Omega_a t
\end{equation}
which yields the FC period $\Omega_{FC}=2\Omega_a$  in the first order in $\lambda$ with
the amplitude $a$ remaining undefined. This amplitude can be determined by the equation $\dot{a}=0$
in Eq. (\ref{deriv}) if the injected amplitude $E_{in}$ is taken as a small parameter $\varepsilon$ in the perturbation approach.
 However that approach is successful only in the fourth order
in $\varepsilon$ resulting in cumbersome equations. Therefore we estimate the
amplitude $a$ averaging the numerical solution over time: $a=\langle a_a(t)\rangle$.
The numerical result shown in Fig. \ref{fig8} (a)
is close to analytical result (\ref{Omega}) in Fig. \ref{fig8} (b) when the injected amplitude is small.
Thus, the period of harmonics generated by the  open nonlinear dimer can be effectively
controlled by the injected amplitude.

Note, the reason for instability related to the bound state in the continuum preserves for $u\neq 0$
and what is more surprising even for different couplings $\gamma_j, j=1, 2$ of the sites
with the injected wave. For this case the CMT equations (\ref{CMTtem}) will take the
following form \cite{Suh}
\begin{eqnarray}\label{CMTdif}
&-i\dot{a_1}=(\nu+\lambda|a_1|^2)a_1+i\gamma_1a_1+i\sqrt{\gamma_1\gamma_2}a_2-i\sqrt{\gamma_1}E_{in},&\nonumber\\
&-i\dot{a_2}=(\nu+\lambda|a_2|^2)a_2+i\sqrt{\gamma_1\gamma_2}a_1+i\gamma_2a_2-i\sqrt{\gamma_2}E_{in}.&
\end{eqnarray}
By linear transformation
\begin{equation}\label{transfor}
    a_1=\sqrt{\gamma_1}(a_s+a_a), ~~a_2=\sqrt{\gamma_2}(a_s-a_a)
\end{equation}
Eqs. (\ref{CMTdif}) take the following form
\begin{eqnarray}\label{asaadif}
&-i\dot{a}_s=\nu a_s+\gamma_+\lambda[(|a_s|^2+2|a_a|^2)a_s+a_a^2a_s^*]+
\gamma_-\lambda[(|a_a|^2+2|a_s|^2)a_a+a_s^2a_a^*]+2i(\gamma_+a_s+\gamma_-a_a)-iE_{in},&\nonumber\\
&-i\dot{a}_a=\nu a_a+\gamma_-\lambda[(|a_s|^2+2|a_a|^2)a_s+a_s^2a_s^*]+
\gamma_+\lambda[(|a_a|^2+2|a_s|^2)a_a+a_a^2a_s^*+\gamma_+a_s^2a_a^*]&
\end{eqnarray}
where $\gamma_{\pm}=(\gamma_1\pm\gamma_2)/2$. One can see that similar to the former symmetric case
$\gamma_1=\gamma_2$ the mode $a_a$ is coupled with the
injected wave only through the nonlinear terms.
\section{summary and discussion}

In this paper we considered one of the simplest nonlinear open system, dimer
whose closed counterpart is exactly integrable system \cite{Eilbeck}.
The term "open" means that a linear waveguide is attached to
the dimer to allow probing the dynamical properties of the dimer.
Even in the case of decoupled ($u=0$) nonlinear sites they interact with each other
through the continuum of the waveguide.
In the framework of coupled mode theory we examined the stability of
stationary solutions of Eq. (\ref{CMTtemp}) in the parametric space of
frequency and amplitude of the probing wave. We found
a domain where stable stationary solutions do not exist. First such domains were found
in open nonlinear plaquette \cite{Maksimov} together with the associated effect of
frequency comb generation. In the present paper we showed a
similar effect for scattering of a monochromatic wave by a nonlinear dimer.

The instability of the open nonlinear dimer is related to a
symmetry protected  bound state in the continuum.
When the dimer is linear there were two eigenmodes, symmetric and antisymmetric.
The symmetrical design of opening of the dimer (see Fig. \ref{fig1}) implies that the injected wave
couples only with the symmetric mode  while there is no direct coupling of the injected wave with
the antisymmetric mode as seen from Eqs. (\ref{asaa}). However owing to nonlinear terms in these equations
the antisymmetric mode $a_a$ is coupled with injected wave through the symmetric mode $a_s$.
Therefore the antisymmetric mode emerges in the response in the vicinity of
the resonance $\nu=0$.

Numerical solution of the temporal coupled mode theory equations (\ref{asaa}) demonstrates
highly nonlinear behavior of the site amplitudes cardinally different from the dynamical
behavior of the closed dimer in the instability domain. Time dependence of these amplitudes
holds many harmonics whose frequencies are equidistantly spaced with the interval  $\Omega$.
This interval which defines the FC period $\Omega_{FC}=2\Omega$ was computed numerically and evaluated
 by the use of asymptotic methods \cite{Bogol} to demonstrate an agreement as shown in
 Fig. \ref{fig8}.
 Respectively, the injected wave after scattering by the nonlinear dimer
acquires these harmonics. The value $\Omega$ goes down with decreasing of the
injected amplitude that opens a way of all-optical control of the harmonics.

\acknowledgments{The work was supported by
RFBR grant 03-02-00497. A.S deeply acknowledges fruitful discussions the problem on frequency comb with
Lushuai S. Cao. The authors also thank D.N. Maksimov and V.V. Val'kov.}

\end{document}